\documentclass[12pt]{iopart}

\usepackage{iopams}
\usepackage{cite}
\usepackage{breqn}
\usepackage{graphicx}
\usepackage{float}
\begin{document}

\title[A paraxial cloak with four lenses of different focal lengths]{A paraxial cloak with four lenses of different focal lengths}

\author{M Revilla, J C Lorenzo and N Hermosa}

\address{National Institute of Physics, University of the Philippines, Diliman, Quezon City, Philippines 1101}
\ead{mrevilla@nip.upd.edu.ph}

\begin{abstract}
We demonstrate cloaking with four lenses of different focal lengths. To achieve this, we compute the separation distances between lenses such that the effective ABCD matrix is equal to just a propagation in free space. Previously, calculations using this method are restricted to 2 pairs of lenses with equal focal lengths. Our computations, on the other hand, had no such restrictions: we generalize even with an asymmetric case. We derive expressions that show the dependence of the distances on the lenses and their focal lengths. Furthermore, we propose a way to minimize the length of this cloak. Our equations reduce to the Rochester cloak when its restriction is imposed. A general paraxial cloak could be useful in advanced undergraduate experiments in optics because: 1) the limit imposed by the focal length has been lifted; and 2) there is a pedagogical benefit of doing experiments with lenses beyond the usual imaging experiments such as reinforcing concepts in Gaussian optics.
\end{abstract}

\vspace{2pc}
\submitto{\EJP}
\maketitle

\section{Introduction}
Invisibility has been a popular idea in science fiction and in photonics research. The general idea behind the perfect cloak is to bend electromagnetic waves around a region, called the cloaking region, while keeping the amplitude and phase of outgoing light preserved. Perfect cloaks have to be invisible in all directions and it should be invisible itself. Although complete omnidirectional invisibility has been ruled out \cite{InvisibleBodies}, many have not dismissed the possibility of making a near-perfect cloak. As a matter of fact, different methods have been studied in the last 20 years to achieve a good invisibility cloak. 

A common method employed in making a cloak is with the use of transformation optics, a technique wherein the coordinates are transformed to control the path of light \cite{pendry2012transformation}. Applying this technique, numerous researchers have simulated and fabricated materials that can direct electromagnetic waves around a cloaking region \cite{CornerHiding,UnderCarpet,IncoherentLightCloak,PolygonalCloak,Permitivity}. Cloaks with multiple cloaking regions have been realized\cite{MultipleRegions}. Complex structures, such as split ring resonators, have also been made from existing materials that can act as wave guides and replicate cloaking \cite{MicrowaveMetamaterial,HomogeneousSilicon,InfraredSplit}. Some of these nano-structures have even succeeded in making polarization-independent cloaks \cite{Polarization}. Moreover, materials with graded index of refraction have been employed to mimic cloaking\cite{3Dgraded,BaTiO3}. This transformation technique has been implemented with terahertz \cite{TerahertzCloak} and radar \cite{RadarCloak,Radar2}, under diffuse background lighting \cite{DiffusiveCloak}, and with unique and arbitrary cloaking regions \cite{CoordTrans,LimitationCloak}. The use of transformation optics has been extended to
other similar phenomena such as acoustics cloaking \cite{AcousticCloak}, heat redirection \cite{HeatFlux}, and magnetic field manipulation \cite{Antimagnets}.

Transformation optics thru geometric optics is perhaps the simplest approach to make a cloak. Over the years, cloaking has been achieved with various optical elements such as mirrors and Fresnel and diverging lenses \cite{CloakLargeObj}, as well as using the refraction of light by using L-shaped glass containers filled with water \cite{WaterCloak} and with the concept of mirage \cite{Mirage}. Interestingly, cloaks using geometrical optics have also been shown to preserve both wave phase and amplitude \cite{wavephasepreservation}. There have also been efforts to look into the spectrum of the incident light and to preserve it by shifting energies to regions that the cloaked object transmits \cite{SpectralCloak}.

By far the most popular among cloaks using geometrical optics is the Rochester cloak made by Howell and Choi, \cite{Rochester,ParaxialPerfect}. Their method makes use of four converging lenses to achieve a perfect paraxial cloak. A \textit{paraxial perfect cloak} preserves, more or less, the incoming field, satisfying some characteristics of a perfect cloak but has only a certain region to hide an object. They are not exempt from edge effects and aberrations. Moreover, the observation is limited to small angles from the optical axis.

In the work of Howell and Choi, they were able to show that one, two, and three lens systems cannot achieve perfect paraxial cloak. They did this by setting the effective ray transfer matrix of the system to correspond to that of free space. They found that with less than 4 lenses, no solutions can be obtained for the separation distances that will satisfy the free space correspondence and that are physically possible. With four lenses, however, solutions for the separation distances were obtained. They were able to check that the cloak indeed has a cloaking region. The limitation in their calculations, however, is that they set a condition that the focal lengths of first and fourth lenses be equal and the other two lenses also have equal focal lengths. They also set the separation distance between first and second lenses equal to that of the third and fourth. With these conditions, their system effectively became two 4$f$ systems, which preserved the input wavefront. Imposing these conditions on the focal length also led to a simpler effective ray transfer matrix, in which the remaining separation distance can be easily solved.

Without imposing conditions on the focal lengths and separation distances, this paper addresses the question: \textit{can cloaking still occur on the four lens system using separation distances solved directly from its effective ray transfer matrix?} We also optimize the system by calculating possible placement of lenses which produce the least total length. With this generalized four lens system, we also test its solution when Rochester cloak conditions are imposed. Lastly, we comment on the applicability of this setup as an undergraduate experiment in optics.

\section{Method}
In making the Rochester cloak, Howell and Choi utilized two ray transfer matrices: that of free space and of a thin lens. The transformations caused by the optical elements under paraxial approximation can be best summarized in terms of ABCD or ray transfer matrices. A system composed of $n$ thin lens and free space units has an effective ABCD matrix that can be expressed as

\begin{equation}
\textbf{M}_{\textnormal{eff}}=\left(\begin{array}{cc}1& 0 \\ -\frac{1}{f_n} &1\end{array}\right)
\prod_{i=1}^{n-1}
\left(\begin{array}{cc}1-\frac{d_{n-i}}{f_{n-i}}& d_{n-i} \\ -\frac{1}{f_{n-i}} &1\end{array}\right)\label{eq:eqn1}
\end{equation}

\noindent where $n$ is the number of lens, $f_n$ is the last lens, $d_i$ is the $i$th separation distance between lens $i$ and $i+1$ and $f_i$ is the focal length of lens $i$. Considering the location of the observer, the $n$th distance, is not necessary in the effective ABCD matrix because the output wavefront is assumed to be preserved after passing through the system. This is also the case for the distance from the source to the first lens. The matrices in the product in equation \ref{eq:eqn1} are the product of a free space and lens ABCD matrices given by\cite{Optics}

\begin{equation}
\left(\begin{array}{cc}1& d \\ 0 &1\end{array}\right)\label{eq:eqn2},
\end{equation}

\noindent for a length of free space $d$, and

\begin{equation}
\left(\begin{array}{cc}1& 0 \\ -\frac{1}{f} &1\end{array}\right),\label{eq:eqn3}
\end{equation}

\noindent for a lens with focal length $f$, respectively. The matrix in equation \ref{eq:eqn1} can be reduced to a 2x2 matrix with elements

\begin{equation}
\textbf{M}_{\textnormal{eff}}=\left(\begin{array}{cc}A_{\textnormal{eff}}& B_{\textnormal{eff}} \\ C_{\textnormal{eff}} &D_{\textnormal{eff}}\end{array}\right)\label{eq:eqn4}
\end{equation}

\noindent where the eff subscript means effective and $A$, $B$, $C$, and $D$ are the elements of the matrices.

Imposing the free space condition on equation \ref{eq:eqn4} means that each elements would have to be equal to that of free space. Hence,

\begin{eqnarray}
A_{\textnormal{eff}}(d_i, d_{i+1}, d_{i+2})=1\label{eq:eqA}\\
B_{\textnormal{eff}}(d_i, d_{i+1}, d_{i+2})=\sum_{i=1}^{n-1}d_i=d_{\textnormal{total}}\label{eq:eqB}\\
C_{\textnormal{eff}}(d_i, d_{i+1}, d_{i+2})=0\label{eq:eqC}\\
D_{\textnormal{eff}}(d_i, d_{i+1}, d_{i+2})=1\label{eq:eqD}.
\end{eqnarray}

\begin{figure}[H]
\centering
\includegraphics[width = 0.90\linewidth]{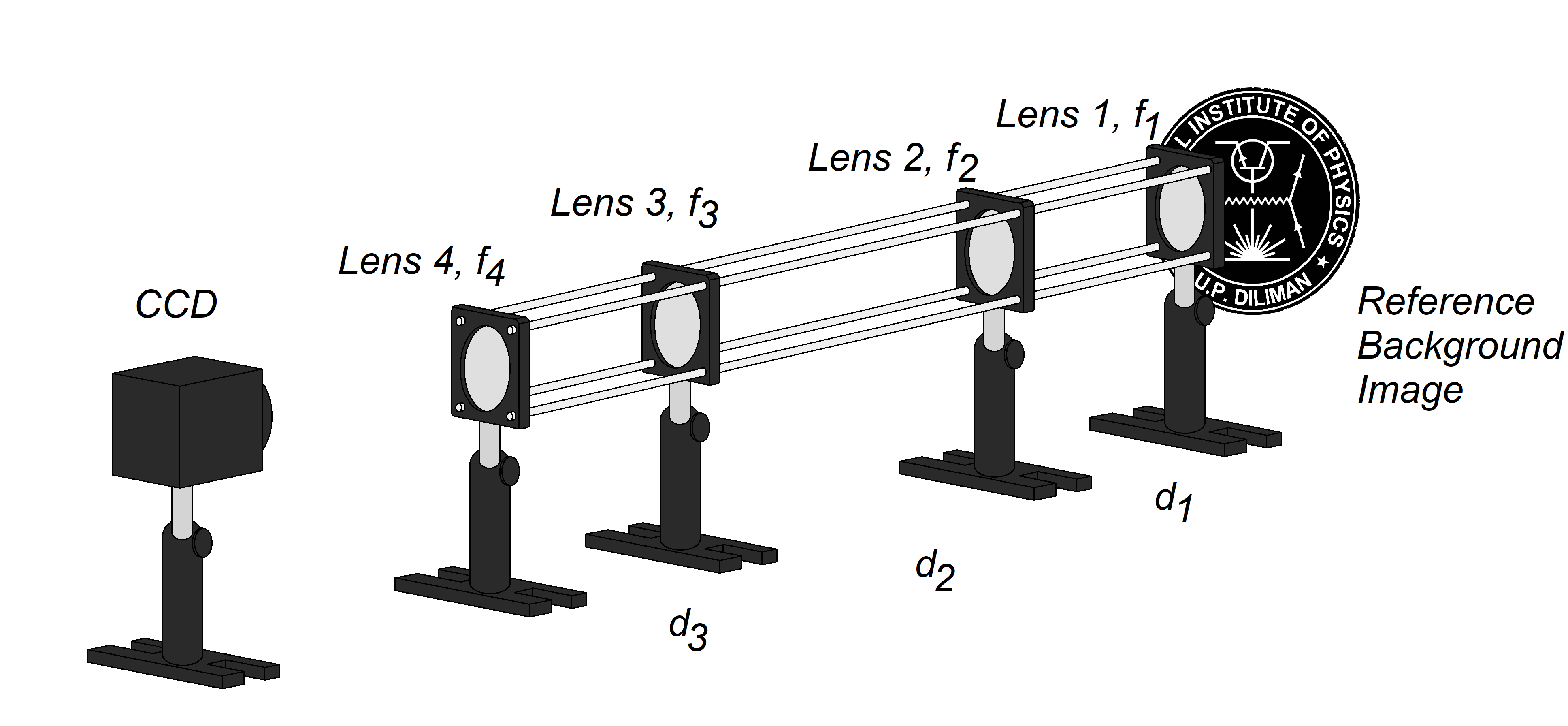}
\caption{The four-lens system. Distances between the lenses are labelled as $d_1, d_2,$ and $d_3$. The focal lengths of lenses 1 to 4 are labelled as $f_1, f_2, f_3,$ and $f_4$ correspondingly.}\label{fig:fig1}
\end{figure}

These constraints are satisfied by the appropriate lens separation distances which can be solved with any combination of the equations above. Equation \ref{eq:eqn1} can be expanded as,

\begin{eqnarray}
\textbf{M}_{\textnormal{eff}}=\left(\begin{array}{cc}1& 0 \\ -\frac{1}{f_4} & 1\end{array}\right)
\left(\begin{array}{cc}1-\frac{d_3}{f_3}& d_3 \\ -\frac{1}{f_3} &1\end{array}\right)
\left(\begin{array}{cc}1-\frac{d_2}{f_2}& d_2 \\ -\frac{1}{f_2} & 1\end{array}\right)
\left(\begin{array}{cc}1-\frac{d_1}{f_1}& d_1 \\ -\frac{1}{f_1} &1\end{array}\right)\label{eq:eff}
\end{eqnarray}

\noindent with the lens system shown in figure \ref{fig:fig1}. 

After imposing free space propagation, Eqs. \ref{eq:eqA}, \ref{eq:eqB}, \ref{eq:eqC} and \ref{eq:eqD} for the 4-lens system will be,

\begin{eqnarray}
A_{\textnormal{eff}}&= \prod_{i=1}^{3}\left(1-\frac{d_i}{f_i}\right) -\frac{d_2}{f_1}\left(1-\frac{d_{3}}{f_{3}}\right) - \frac{d_3}{f_2}\left(1-\frac{d_1}{f_1}\right) - \frac{d_3}{f_1}=1 \label{eq:eqA1}\\
B_{\textnormal{eff}}&= d_1\prod_{i=1}^{2}\left(1-\frac{d_{i+1}}{f_{i+1}}\right)+d_2\left(1-\frac{d_3}{f_3}\right)+d_3\left(1-\frac{d_1}{f_2}\right)=d_{\textnormal{total}}
\label{eq:eqB1}\\
C_{\textnormal{eff}}&= -\frac{1}{f_4}A_{\textnormal{eff}}-\frac{1}{f_3}\prod_{i=1}^{2}\left(1-\frac{d_i}{f_i}\right)+\frac{d_2}{f_3f_1}-\frac{1}{f_2}\left(1-\frac{d_1}{f_1}\right)-\frac{1}{f_1}=0 \label{eq:eqC1}\\
D_{\textnormal{eff}}&=-\frac{1}{f_4}B_{\textnormal{eff}}-\frac{d_1}{f_3}\left(1-\frac{d_2}{f_2}\right)+\left(1-\frac{d_1}{f_2}\right)-\frac{d_2}{f_3}=1 \label{eq:eqD1}
\end{eqnarray}

\noindent Any 3 of these equations and knowing that $d_{\textnormal{total}}=d_1+d_2+d_3$, the distances $d_i$'s can be solved.

Technically, the number of lenses beyond 4 may be used for cloaking. However, additional constraints must be imposed to match the number of unknowns. Moreover,	realizing a more than 4-lens system in experiments can be challenging as it is very sensitive on the alignment of the lenses.

\section{Results and Discussion}
The ensuing system of equations, Eqs. \ref{eq:eqA1} to \ref{eq:eqC1} and $d_{\textnormal{total}}$, solutions for $d_1$, $d_2$, and $d_3$ are given by,

\begin{equation} \label{eq:eqnd1}
d_1 = \alpha \frac{1 - \sqrt{\frac{f_1f_2}{f_3f_4}}}{f_3f_4 - f_1f_2}
\end{equation}

\begin{equation} \label{eq:eqnd2}
d_2 = \alpha \frac{1 + \sqrt{\frac{f_2f_3}{f_1f_4}}}{f_1f_4 - f_2f_3}
\end{equation}

\begin{equation} \label{eq:eqnd3}
d_3 = \alpha \frac{1 - \sqrt{\frac{f_3f_4}{f_1f_2}}}{f_1f_2 - f_3f_4}
\end{equation}

\noindent where
\begin{equation}
\alpha=f_2f_3\left(f_1+f_4\right)+f_1f_4\left(f_2+f_3\right)\label{eq:eqnalpha}
\end{equation}

These separation distances, Eqs. \ref{eq:eqnd1} to \ref{eq:eqnalpha}, are the main results of this paper. The $d_i$'s are functions of the focal lengths only. These illustrate that relaxing most of the conditions set in the Rochester cloak can still yield viable, or realizable separation distances given appropriate choice of focal lengths. These equations, however, must still hold true if ever focal length restrictions of the Rochester cloak were to be imposed. We show that it is indeed so, later in our discussion. 

\begin{figure}[H]
\centering
\includegraphics[width=0.55\textwidth]{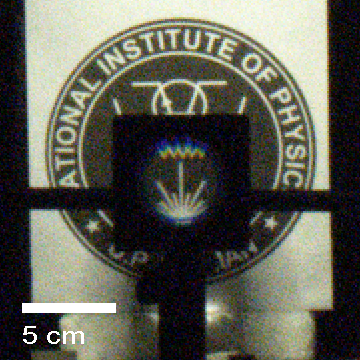}
\caption{The cloak in action as viewed on-axis. A metal rod is placed horizontally inside the cage system. The background, the National Institute of Physics logo, can still be viewed even though an obstruction is placed inside. Lenses 1-4 for this system had focal lengths of 20.0 cm, 6.0 cm, 7.5 cm, and 17.5 cm, respectively. Chromatic aberrations can be seen at the edge of the lens.}\label{fig:fig2}
\end{figure}

As proof of principle, a cloak was designed using four 2'' uncoated N-BK7 bi-convex lenses from ThorLabs. The focal length of these lenses were 20.0 cm (lens 1), 6.0 cm (lens 2), 7.5 cm (lens 3), and 17.5 cm (lens 4). Using equations \ref{eq:eqnd1} to \ref{eq:eqnalpha}, the separation distances given these focal lengths were $d_1 = 25.0$ cm, $d_2 = 28.6$ cm, and $d_3= 26.1$ cm, for a total length of 79.7 cm. A logo of the institute was used as background image. The object was placed between lenses 3 and 4 and its position inside was adjusted to observe the cloaking phenomenon. The images were taken by a PointGrey Grasshopper 3 4.1 megapixel CCD camera with a Tamron 1/1.8'' imaging lens. Although, any camera with a lens will do. 

\begin{figure}[H]
\centering
\includegraphics[width=0.55\textwidth]{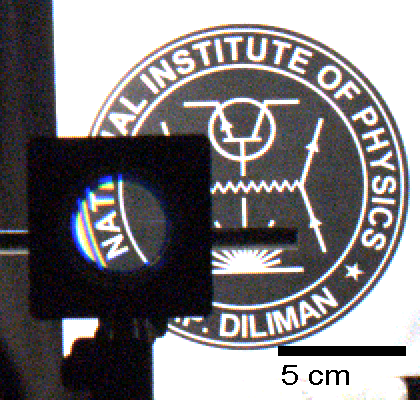}
\caption{The cloak with the background image off-centred. By placing the background as such, magnification and image distortion can easily be checked by looking at the circular edge of the logo. The lenses used in this system were the same as those in figure \ref{fig:fig2}.}\label{fig:fig3}
\end{figure}

Similar to the Rochester cloak, cloaking happens when the object is placed off axis from the centre: it does not happen when the object is placed at the centre of the lens system. Moreover, cloaking can be perceived only at a small angle with respect to the optical axis similar to  the Rochester cloak. Since we have used uncoated non-achromatic lenses, chromatic aberrations are present at the edges of the lens. These aberrations are also seen in the image as some brightly coloured bands at the edge of the cloak. Furthermore some magnification is observed in the background image.  This magnification and aberration are further highlighted in figure \ref{fig:fig3}. However, \emph{we believe that we clearly demonstrate cloaking with our setup.}

One particular interest in this system is how to minimize its total length $d_{\textnormal{total}}$. With Eqs. \ref{eq:eqnd1} to \ref{eq:eqnalpha}, $d_{\textnormal{total}}$ can be expressed as,

\begin{eqnarray}
d_{\textnormal{total}}&=\alpha\left(\frac{1-\sqrt[]{\frac{f_1f_2}{f_3f_4}}}{f_3f_4-f_1f_2}+\frac{1+\sqrt[]{\frac{f_2f_3}{f_1f_4}}}{f_1f_4-f_2f_3}-\frac{1-\sqrt[]{\frac{f_3f_4}{f_1f_2}}}{f_3f_4-f_1f_2}\right)\\
&=\alpha\left(\frac{\sqrt[]{\frac{f_3f_4}{f_1f_2}}-\sqrt[]{\frac{f_1f_2}{f_3f_4}}}{f_3f_4-f_1f_2}+\frac{1+\sqrt[]{\frac{f_2f_3}{f_1f_4}}}{f_1f_4-f_2f_3}\right)\label{eq:total}
\end{eqnarray}

Looking at Eq.\ref{eq:total}, the first term has a difference for its numerator, meanwhile, the second term has a sum. A way to minimize $d_{\textnormal{total}}$ is to make the second term of the sum smaller while keeping the first term relatively suppressed. This can be achieved by greatly increasing the difference $f_1f_4-f_2f_3$, denoted as $\Delta_{1423}$. By making values of $f_1$ and $f_4$ close and considerably larger than those of $f_2$ and $f_3$, $\Delta_{1423}$ is increased while keeping the first term roughly the same.

\begin{figure}[H]
\centering
\includegraphics[width=0.7\textwidth]{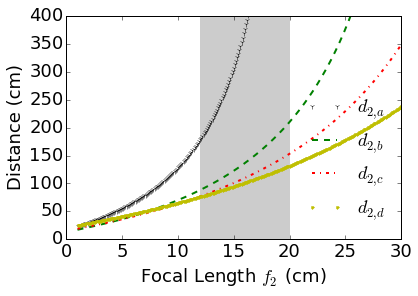}
\caption{Plot of $d_2$ values for increasing $f_2$, effectively decreasing $\Delta_{1423}$. In $d_{2,a}$, the values of $f_1$, $f_3$, and $f_4$ were 18 cm, 16 cm, 20 cm, respectively. As for $d_{2,b}$: 20 cm, 12 cm, and 22 cm; for $d_{2_c}$: 25 cm, 14 cm, and 27.5 cm; and for $d_{2_d}$: 35 cm, 20 cm, 40 cm. The shaded region in the figure points where $f_2$ ranges closely with its corresponding $f_3$ value.}\label{fig:fig4}
\end{figure}

Figure \ref{fig:fig4} illustrates this idea of increasing $\Delta_{1423}$ to diminish $d_2$, and in turn $d_{\textnormal{total}}$, to workable distances. Shown in the figure are different set-ups of a general four-lens cloak, labelled from $a$ to $d$. Comparing the behaviour of $d_2$ in set $a$ (where $f_1$, $f_3$, and $f_4$ are 18 cm, 16 cm, and 20 cm) to its value in set $d$ ($f_1$, $f_3$ and $f_4$: 35 cm, 20 cm, 40 cm), there is a noticeable difference between their $d_2$ values: $\approx$ 200 cm for set $a$ and $\approx$ 140 cm for set $d$.

Lastly, we looked into equations \ref{eq:eqnd1} to \ref{eq:eqnalpha} under Rochester cloak-like conditions: $f_1=f_4$ and $f_2=f_3$. The equations for the separation distances transform into

\begin{equation}
d_1^{\textnormal{R}}=\lim_{f_1\to f_4, f_2\to f_3}\alpha\frac{1-\sqrt[]{\frac{f_1f_2}{f_3f_4}}}{f_3f_4\left(1-\frac{f_1f_2}{f_3f_4}\right)}=f_1+f_2\label{eq:d1mod}
\end{equation}

\begin{equation}
d_2^{\textnormal{R}}=\lim_{f_1\to f_4, f_2\to f_3}\alpha\frac{1 + \sqrt{\frac{f_2f_3}{f_1f_4}}}{f_1f_4 - f_2f_3}=\frac{2f_2(f_1+f_2)}{f_1-f_2}\label{eq:d2mod}
\end{equation}

\begin{equation}
d_3^{\textnormal{R}}=\lim_{f_2\to f_3, f_4\to f_1}\alpha\frac{1-\sqrt[]{\frac{f_3f_4}{f_1f_2}}}{f_1f_2\left(1-\frac{f_1f_2}{f_3f_4}\right)}=f_3+f_4=f_1+f_2\label{eq:d3mod}
\end{equation}

With Eqs. \ref{eq:d1mod} and \ref{eq:d3mod}, $d_1$ reduces to $f_1+f_2$ and $d_3$ to $f_3+f_4$ which is equal to $f_1+f_2$ under Rochester cloak conditions. Such result agrees with the findings of Howell and Choi. In addition, $d_2$ simplifies into $2f_2(f_1+f_2)/(f_1-f_2)$, which is also consistent with the Rochester cloak. Equations for separation distances in the Rochester cloak can be recovered upon application of the necessary $f$ to the formulas we presented. This fact provides proof that our equations are consistent with earlier findings.

\section{Conclusions}
We made an optical cloak with four lenses of different focal lengths by determining appropriate separation distances in terms of the lenses' focal lengths. In addition, we propose a criterion for minimizing the total length  by making both $f_2$ and $f_3$ smaller compared to the values of $f_1$ and $f_4$, given that the values of $f_2$ and $f_3$ as well as those of $f_1$ and $f_4$ are close together. The equations for the separation distances we presented also reduce to the Rochester cloak calculations with the appropriate focal length conditions.

Given the cloak's straightforward calculations and method, it can be included in undergraduate optics experiments. Students may be asked to setup and solve the solution for the distances with Eqns. \ref{eq:eqnd1} to \ref{eq:eqnalpha} together with the expression for $d_{\textnormal{total}}$. This can serve as an introduction for physics undergraduates to analytical ray tracing method via the ABCD matrices. This is beyond the usual image formation computations that are currently used in undergraduate laboratory classes. Furthermore with focal length constraints relaxed, the cloaking phenomenon is now more easily replicable. Teachers can readily obtain distances between accessible lenses. This way, cloaking can be displayed on optics exhibits and in-class demonstrations effortlessly. A setup that cloaks is usually a major draw in exhibits. It can be an opportunity to educate and excite young minds. 

\section*{Acknowledgements}
N. Hermosa is a Balik PhD recipient of the University of the Philippines Office of the Vice President for Academic Affairs (Balik PhD 2015-06). This work is funded by the Department of Science and Technology's Grants for Outstanding Achievements in Science and Technology through the National Academy of Science and Technology.

\section*{References}
\bibliographystyle{iopart-num}
\bibliography{iopart-num}

\end{document}